\documentclass[twocolumn]{openjournal}

\usepackage[T1]{fontenc}
\usepackage{ae,aecompl}

\usepackage{graphicx}	
\usepackage{amsmath}	
\usepackage{amssymb}	
\usepackage{natbib}

\usepackage{hyperref}
\hypersetup{
	colorlinks=true,
	urlcolor=blue,    
	linkcolor=black,  
	citecolor=blue,   
}

\defcitealias{Pittordis_2018}{PS18}
\defcitealias{Pittordis_2019}{PS19}
\defcitealias{Badry_2021}{ERH}
\defcitealias{Banik_2018_Centauri}{BZ18}

\hypersetup{citecolor=blue, 
            linkcolor=red, 
            menucolor=blue, 
            urlcolor=blue}  

\newcommand{\cm}{{~\rm cm}}

\newcommand{\km}{{~\rm km}}
\newcommand{\s}{{~\rm s}}

\newcommand{\g}{{~\rm g}}

\newcommand{\yr}{{~\rm yr}}

\newcommand{\kpc}{{~\rm kpc}}

\begin{document}

\title{Bright common envelope formation requires jets}


\author{Noam Soker} 
\affiliation{Department of Physics, Technion, Haifa, 3200003, Israel;  soker@physics.technion.ac.il}

\begin{abstract}
I compared with each other and with observations three energy sources to power intermediate luminosity optical transients (ILOTs) and conclude that only jets can power bright ILOTs with rapidly rising lightcurves. I present an expression for the power of the jets that a main sequence secondary star launches as it enters a common envelope evolution (CEE) with a primary giant star. The expression includes the Keplerian orbital period on the surface of the primary star, its total envelope mass, and the ratio of masses. I show that the shock that the secondary star excites in the envelope of the primary star cannot explain bright peaks in the lightcurves of ILOTs, and that powering by jets does much better in accounting for rapidly rising, about 10 days and less, peaks in the lightcurves of ILOTs than the recombination energy of the ejected mass.
I strengthen previous claims that jets powered the Great Eruption of Eta Carinae, which was a luminous variable major eruption, and the luminous red novae (LRNe) V838 Mon and V1309 Scorpii.  I therefore predict that the ejecta (nebula) of V1309 Scorpii will be observed in a decade or two to be bipolar. My main conclusion  is that only jets can power a bright peak with a short rising time of ILOTs (LRNe) at CEE formation. 
\end{abstract}

\keywords{stars: AGB and post-AGB; stars: jets; stars: mass-loss;  stars: variables: general; binaries (including multiple): close; planetary nebulae: general}

\section{Introduction}
\label{sec:intro}

Intermediate luminosity optical transients (ILOTs) are gravitationally-powered transient events with peak luminosities ranging from somewhat below peak luminosities of classical novae and up to those of typical supernovae  (e.g., \citealt{Mouldetal1990, Rau2007, Ofek2008, Masonetal2010, Kasliwal2011, Tylendaetal2011, Tylendaetal2013, Kasliwal2013, Tylendaetal2015, Kaminskietal2018, Pastorelloetal2018, BoianGroh2019, Caietal2019, Jencsonetal2019, PastorelloMasonetal2019, Banerjeeetal2020, Stritzingeretal2020b, Blagorodnovaetal2021, Pastorelloetal2021, Bondetal2022,
Caietal2022a, Caietal2022b, Wadhwaetal2022, Karambelkaretal2023, Pastorelloetal2023}). The release of gravitational energy can be by mass transfer in a binary system, most likely with the launching of jets by the mass-accreting star (e.g., \citealt{Kashietal2010, SokerKashi2016TwoI, Kashi2018Galax, Soker2020ILOTjets}), and/or mass ejection in the equatorial plane (e.g., \citealt{Pejchaetal2017, HubovaPejcha2019}). Alternatively, the gravitational energy might result from a merger process of two objects, including the onset of a common envelope evolution (CEE). The later is the focus of this study. 

When spatially resolved, the nebulae that ILOTs form are bipolar. The Homunculus nebula that was formed during the Great Eruption of Eta Carinae (e.g., \citealt{DavidsonHumphreys1997}) is bipolar. This is a luminous blue variable major eruption which is a type of ILOT.  The ILOT V4332~Sgr ejected a bipolar nebula \citep{Kaminskietal2018}. Most interesting is the ejecta of  
Nova~1670 (CK~Vulpeculae) with its 350-years old bipolar nebula \citep{Sharaetal1985} because it has an S-morphology \citep{Kaminskietal2020Nova1670, Kaminskietal2021CKVul}. An S-morphology suggests shaping by precessing ejts. 

I will use the term luminous red novae (LRNe) for those ILOTs that are powered by full merger, but are fainter than typical supernovae. This includes also CEE, whether the companion survives or not the CEE\footnote{There is no consensus on these terms (see also \citealt{Caietal2022b} for a review). I refer to all gravitationally-powered transient events as ILOTs (for the usage of the term ILOT see, e.g., \citealt{Berger2009, KashiSoker2016, MuthukrishnaetalM2019}). Some do not use the ILOT term (e.g., \citealt{Jencsonetal2019}). There is also no consensus on the set of sub-classes, e.g., \cite{KashiSoker2016} versus  \cite{PastorelloFraser2019} and \cite{PastorelloMasonetal2019}.}.  
When a neutron star or a black hole spiral-in inside the envelope of a red supergiant (RSG) star and launch jets the event can be very bright as to mimic a core collapse supernova (CCSN), or even a superluminous CCSN (e.g., \citealt{SokerGilkis2018, Gilkisetal2019, GrichenerSoker2019, YalinewichMatzner2019, Schreieretal2021}). These events are termed common envelope jets supernovae (CEJSNe). 

LRNe can result from the onset of a CEE with a stellar companion (e.g., \citealt{Tylendaetal2011, Ivanovaetal2013a, Nandezetal2014, Kaminskietal2015, Pejchaetal2016a, Pejchaetal2016b, Soker2016GEE, Blagorodnovaetal2017, MacLeodetal2017, MacLeodetal2018,  Segevetal2019, Howittetal2020, MacLeodLoeb2020, Qianetal2020, Schrderetal2020, Blagorodnovaetal2021, Addison2022, MatsumotoMetzger2022, Zhuetal2023}) or with a sub-stellar (i.e., a brown dwarf or a planet) companion (e.g., \citealt{RetterMarom2003, Retteretal2006, Metzgeretal2012, Yamazakietal2017, Kashietal2019Galax, Gurevichetal2022, Deetal2023, Oconnoretal2023}) 

\cite{Ivanovaetal2013a} study the emission at CEE formation by the recombination of the ionized ejecta (also \citealt{Howittetal2020}). They consider a spherical ejecta, which we now know is not the case with resolves ILOTs because they have bipolar ejecta. This model cannot explain all properties, including the rapid, few days, rise to the second peak (first bright peak) in the light curve of V838Mon and the total energy of the Great Eruption of Eta Carinae. In section \ref{sec:accretion} I argue that accretion energy that jets deliver is more efficient than recombination energy. Recombination can nicely explain the plateau phase of lightcurves. \cite{MatsumotoMetzger2022} conduct detailed calculations of the lightcurves due to recombination in CEE. Their results show that the recombination cannot account for a rapidly rising peak and cannot explain some luminous LRNe. For example, in the recombination model of \cite{MatsumotoMetzger2022} the LRN AT~2014ej requires unreasonable high mass ejection (below I mention that \citealt{SokerKaplan2021RAA} could fit this LRN with jets).

I studied the role of jets in powering ILOTs in several earlier papers. 
In \cite{Soker2016GEE} I studied the energetic of jet-powering in the grazing envelope evolution (GEE). I concluded that during the GEE the jets accelerate the outskirts of the primary envelope to velocities that in some cases bring the flow time to be of the order of the photon diffusion time out of the ejecta, $t_{\rm flow} \approx t_{\rm diff,ej}$. This implies that a large fraction of the kinetic energy of the jets, which is channelled to thermal energy as the jets collide with the envelope, ends in radiation. This leads to a bright ILOT event. However, in that study I did not examine the luminosity nor did I calculate the accretion rate onto the companion that enters the GEE with the more extended primary star.   

In \cite{Soker2020ILOTjets} I demonstrated that the interaction of wide jets with a slower expanding shell is very efficient in channelling the jets' kinetic energy to radiation. I applied this interaction to three ILOTS: the Great Eruption of Eta Carinae (observational data from, e.g., \citealt{DavidsonHumphreys2012Natur, Restetal2012}), to V838 Mon (physical data from \citealt{Tylenda2005}), and to V4332 Sgr (data from, e.g., \citealt{Kaminskietal2018}). I found the jet-shell interaction to be more efficient in converting kinetic energy to radiation than collision of equatorial outflows (e.g., \citealt{Pejchaetal2016a, Pejchaetal2016b}). The collision of a fast spherically ejected shell with a slow dense equatorial gas (e.g., \citealt{AndrewsSmith2018, KurfurstKrticka2019}) is more efficient than collision of equatorial ejecta (e.g., \citealt{MetzgerPejcha2017}), but not as that of jet-shell interaction. In any case, fast spherical shell hitting a slower equatorial outflow is not a flow structure that occurs at the formation of CEE. In \cite{Soker2020ILOTjets} I did not deal with the onset of the CEE.

\cite{SokerKaplan2021RAA} apply toy models of jet-shell interaction to reproduce the light curves of the ILOTs SNhunt120 (observations by \citealt{Stritzingeretal2020a}) and AT 2014ej (an LRN; observations by \citealt{Stritzingeretal2020a}). \cite{SokerKaplan2021RAA} could fit the light curves with jet-shell interaction (note that the units of times in their two figures should be day). \cite{SokerKaplan2021RAA} did not consider the onset of the CEE.   
These three studies of jet-shell interaction did not examine the accretion power of the compact object that launches the jets nor did they compare the power with the collision of the companion with the extended envelope.

In the present study I first (section \ref{sec:Jetless}) show that the shock that the companion excite in the envelope as it enters a CEE cannot lead to a bright ILOT. In section \ref{sec:accretion} I compare the luminosity from jets to that from the recombination energy of the ejecta. In section \ref{sec:Systems} I apply the results to several ILOTs. I summarize in section \ref{sec:Summary}. This study is motivated in part by the recent finding by \cite{Mobeenetal2023} that the innermost circumstellar material currently surrounding V838 Mon is bipolar. 

\section{Powering by jetless interaction}
\label{sec:Jetless}

In CEE, unlike in supernovae, recombination energy releases energy over a relatively long time, larger than the dynamical time as it requires the envelope to expand, cool, recombines, and for the photons to diffuse out (e.g.,  \citealt{MatsumotoMetzger2022}). Recombination powering cannot explain the rapid rise, less than about 10 days, that is observed in many ILOTs, from luminous blue variables to LRNe (e.g., \citealt{Caietal2022b} for a review). 

\begin{figure}[t]
	\centering
\includegraphics[trim=0.15cm 0.2cm 0.0cm 2.0cm ,clip, scale=0.42]{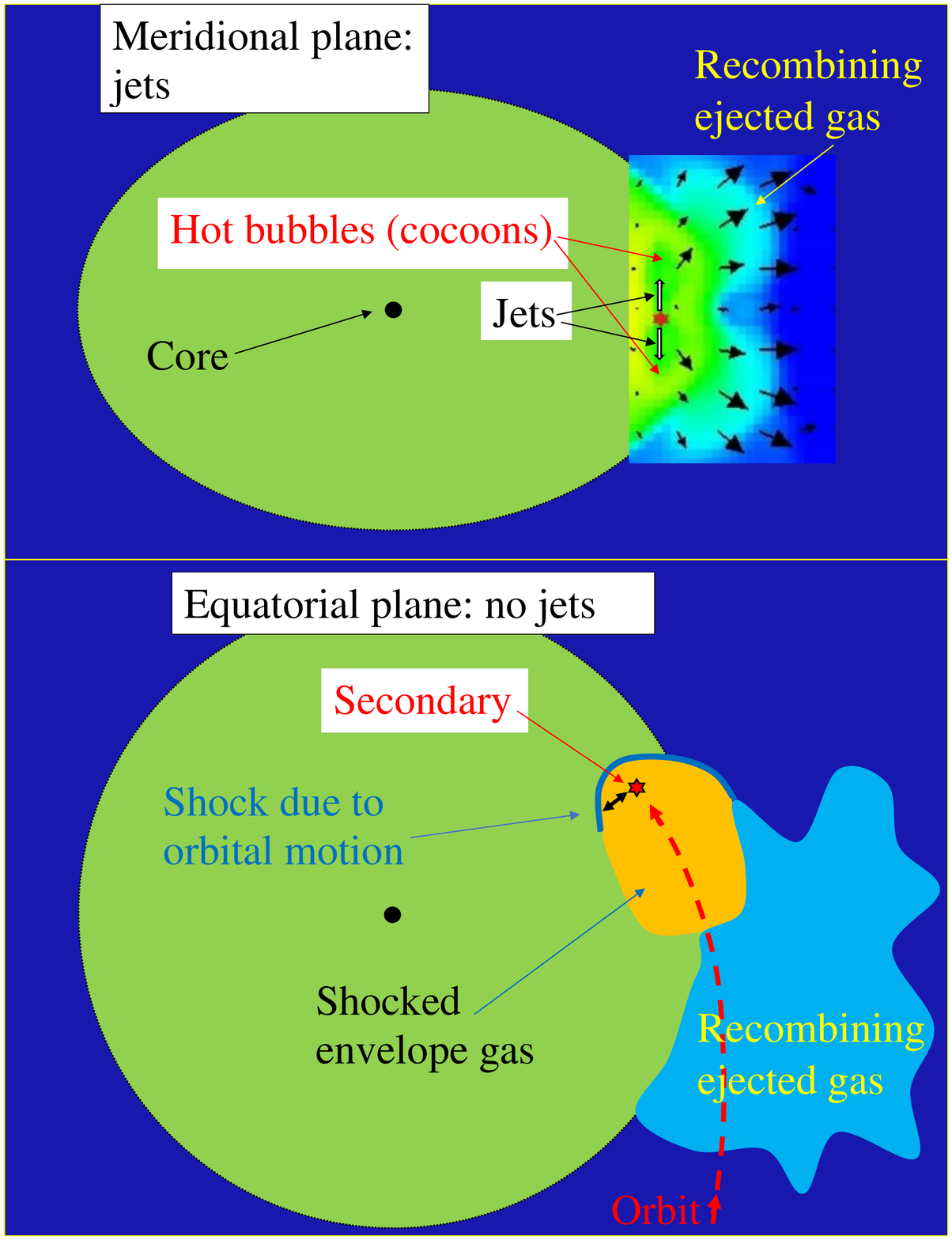} 
\caption{ A schematic drawing (not to scale) of the three discussed energy sources. Dark blue is a very low-density zone around the binary systems, i.e., the wind of the giant star. The green region is the envelope of the giant star, the small black circle is the core of the giant star, and the red star of David is the secondary star that accretes mass. 
Upper panel: A schematic drawing in the meridional plane that momentarily contains both the core and the secondary star. The inset with that black arrows is from figure 2 of \cite{Schreieretal2023} which is a three-dimensional hydrodynamical simulations of a CEE with jets. The black arrows depict the outflow map. The yellow-coloured zones are higher density than the envelope density (green) that are compressed by the bubble (cocoon) that each of the two jets inflates. Both the shock that the secondary star excite in the envelope (not shown in upper panel) and recombination exist in this case.   
Lower panel: A schematic drawing in the equatorial plane when the secondary star does not launch jets and shortly after it entered the envelope of the giant star. The dashed-red line depicts the orbit. The double-headed arrow touching the secondary star is about the length of $R_{\rm BHL}$. 
}
\label{fig:fig1}
\end{figure}

I schematically present the jetless interaction in the lower panel of Fig. \ref{fig:fig1}. The powering includes recombination energy and the shock that the secondary star excites in the envelope.   

A rapid interaction takes place when the secondary star, with mass $M_2$ and a small radius $R_2 \ll a$ where $a$ is the orbital radius, enters the envelope of the primary star, i.e., during the early CEE phase when $a \simeq R_1$. I depict this flow structure in the lower panel of Fig. \ref{fig:fig1}. The primary, of mass $M_1$ and radius $R_1$, can be a main sequence and up to a RSG. The secondary excites a shock wave with a typical distance from the orbit which is about the classical Bondi-Hoyle-Lyttleton radius 
\begin{equation}
R_{\rm BHL}= \frac{2 G M_2}{c^2_s(a) + v^2_{\rm rel}(a)} \simeq  \frac{2 M_2}{M_1}a,  
\label{eq:Rbhl}
\end{equation} 
where $v_{\rm rel} \simeq \sqrt{G M_1/a}-v_{\rm rot}(a)$, $v_{\rm rot}(a)$ is the rotation velocity of the envelope at the location of the secondary star, and $c_s(a)$ is the sound speed. In the second equality I took the denominator to be ${c^2_s(a) + v^2_{\rm rel}(a)} \simeq v^2_{\rm Kep} (a) = G M_1/a$. Namely, I assume that the contributions of the sound speed in the envelope and of the envelope rotation more or less cancel each other. To the accuracy of the present study this approximation is adequate. 

The Power of the shock, i.e., the rate of channelling kinetic energy to thermal energy, depends on the mass inflow rate into the shock $\dot M_{\rm shock}$, its velocity $v_{\rm shock}$, the Mach number, and the shock angle. For an approximate expression in the case of a high mach number shock in the present case I take a perpendicular shock out to radius $R_{\rm BHL}$ and $v_{\rm shock} \simeq v_{\rm rel} \simeq \sqrt{GM_1/a}$, so that $\dot M_{\rm shock} \simeq \pi R^2_{\rm BHL} \rho(a) v_{\rm rel}$. This gives for the shock power  
\begin{eqnarray}
\begin{aligned} 
\dot E_{\rm shock} & \simeq \frac{1}{2} \dot M_{\rm shock} v^2_{\rm shock} \simeq
\frac{\pi}{2} R^2_{\rm BHL} v^3_{\rm rel} \rho
\\ & \simeq 
2 \pi \left( \frac{M_2}{M_1} \right)^2 \left( G M_1 \right)^{3/2} a^{1/2} \rho.
\label{eq:Eshock1}
\end{aligned}
\end{eqnarray}
The density in the envelope of giant stars varies as $\rho(r) \propto r^{-\beta}$ with $\beta \simeq 2-3$. This implies that  as the secondary spirals-in the power of the shock increases. However, deeper in the envelope the photon diffusion time rapidly increases inward, implying that the fraction of the thermal energy that end in radiation decreases.  
The expansion time of the post-shock gas inside the accretion radius under the present assumptions, is $\tau_{\rm exp} \simeq R_{\rm BHL} / v_{\rm shock} \simeq R_{\rm BHL} / v_{\rm Kep} \equiv \tau_{\rm f}$. 
The photon diffusion time is $\tau_{\rm dif,2} \simeq \tau R_{\rm BHL}/c$ where $\tau\simeq \rho \kappa  R_{\rm BHL}$ is the optical depth and $\kappa$ the opacity. I return to these timescales in section \ref{subsec:TimeScales}. I assume that the maximum luminosity is obtained where $\tau_{\rm exp} \simeq  \tau_{\rm dif,2}$. 
This gives for the density at maximum luminosity due to the CEE shock
\begin{equation}
\rho_{\rm max-L} \simeq \frac{c}{\kappa R_{\rm BHL}  v_{\rm Kep}} .
\label{eq:rhoMaxL}
\end{equation} 
Substituting this density in equation (\ref{eq:Eshock1}) and assuming that about half of the shock energy ends in radiation, because $\tau_{\rm dif,2} \simeq \tau_{\rm exp}$, yields 
\begin{eqnarray}
\begin{aligned} 
& L_{\rm shock}  \simeq \dot E_{\rm shock} (\rho_{\rm max-L} ) 
\simeq
\frac{\pi}{2} \frac{c}{\kappa}  GM_2
\\ & \simeq 1.6 \times 10^4 
\left( \frac{\kappa}{0.1 \cm^2 \g^{-1}} \right)^{-1} 
\left( \frac{M_2}{1M_\odot} \right) L_\odot  .
\label{eq:Lrad}
\end{aligned}
\end{eqnarray}
For scattering on electron opacity the value of $\kappa$ is larger, implying in turn lower peak luminosity. 

As discussed later, the expression (\ref{eq:Lrad}) gives shock luminosity which is much fainter than most ILOTs. 

\section{Jet powering}
\label{sec:accretion}

\subsection{The power of the jets}
\label{subsec:JetsPower}
In \cite{Soker2016GEE} I studied the accretion total energy versus the recombination total energy. Here I concentrate on the luminosity. It is useful to express the accretion power at CEE formation in terms of global properties (as masses and orbital period) of the binary system rather than local quantities (i.e., velocity and density). 
I examine the interaction at CEE formation (onset), i.e., $a \simeq R_1$. I schematically present this flow in the upper panel of Fig. \ref{fig:fig1}. I take the envelope density profile as $\rho \propto r^{-\beta}$, where for giant (red giant branch, RGB; asymptotic giant branch, AGB;  RSG) stars $\beta \simeq 2-3$, and so the envelope mass is \begin{equation}
M_{\rm env} = \frac{4 \pi}{3-\beta} \rho (R_1) R^3_1, \qquad \beta <3. 
\label{eq:Menv}
\end{equation} 

I take the power of the two jets that the secondary star launches to be $\dot E_{\rm 2j} = \eta_{\rm 2j} \dot E_{\rm acc}$, where the BHL  accretion power is $E_{\rm acc} = G M_2 \dot M_{\rm BHL} /2R_2$. Using the same assumptions as in section \ref{sec:Jetless}, i.e., equation (\ref{eq:Rbhl}), gives 
for $a \simeq R_1$
\begin{eqnarray}
\begin{aligned} 
\dot E_{\rm 2j} (R_1) & \simeq \eta_{\rm 2j}  \pi \frac{G M_2}{2 R_2} R^2_{\rm BHL} \rho (R_1) v_{\rm rel} (R_1) 
\\ & \simeq  
\left( \frac{M_2}{M_1} \right)^2
\eta_{\rm 2j}  \pi (3-\beta) \frac{G M_2 M_{\rm env}}{R_2 \tau_{\rm Kep}} ,
\label{eq:E2j2}
\end{aligned}
\end{eqnarray}
where $\tau_{\rm Kep} = 2 \pi R^{3/2}_1 / (GM_1)^{1/2}$ is the Keplerian orbital period on the surface of the primary star and where quantities are calculated near the primary surface, i.e., when $a \simeq R_1$. 

I scale equation (\ref{eq:E2j2}) for a main sequence companion to yield 
\begin{eqnarray}
\begin{aligned} 
\dot E_{\rm 2j} (R_1) & \simeq 10^5 
(3-\beta)  \left( \frac{\eta_{\rm 2j}}{0.1} \right)   
\left( \frac{M_2/R_2}{M_\odot/R_\odot} \right)  
\\ & \times 
\left( \frac{M_2}{0.1 M_1} \right)^2
\left( \frac{M_{\rm env}}{1 M_\odot} \right)
\left( \frac{\tau_{\rm Kep}}{1 \yr} \right)^{-1} L_\odot .
\label{eq:E2j3}
\end{aligned}
\end{eqnarray}

In the derivation here the parameter $\eta_{\rm 2j}$ includes both the reduction in the actual accretion rate relative to the BHL accretion rate, and the fraction of energy carried by the jets relative to the actual accretion energy (see also section \ref{subsec:TimeScales}).

If a fully ionized solar-composition gas is ejected from the envelope and recombines on a rate of  $\dot M_{\rm ej,rec}$ as it leaves the star and cools, the power the recombination releases is 
\begin{equation}
\dot E_{\rm rec} = 2.5 \times 10^5  
\left( \frac{\dot M_{\rm ej,rec}}{1 M_\odot \yr ^{-1}} \right)
L_\odot .
\label{eq:Erec}
\end{equation}
This is actually an upper limit on the radiation due to recombination because a fraction of this energy might end in unbinding envelope gas and accelerating it out.  

Numerical simulations of CEE that do not include extra energy source of recombination (which here is assumed to end in radiation) and jets show that the ejection of the envelope proceeds on a longer time than the Keplerian orbital period on the surface of the star (as I discuss in section \ref{subsec:PNe}). A timescale that is not much longer than several times the Keplerian period on the surface of the primary requires a secondary of mass $M_2 > 0.1 M_1$. 
The timescale over which the recombination power is released is longer than the ejection time itself, i.e., the time to unbind the gas, because the gas needs to expand and cool for it to recombine.   
From equations (\ref{eq:E2j3}) and (\ref{eq:Erec}) I scale therefore the power ratio of the two energy sources to be  
\begin{eqnarray}
\begin{aligned} 
\frac{\dot E_{\rm 2j} (R_1)}{\dot E_{\rm rec}} &  \simeq 4
(3-\beta)  \left( \frac{\eta_{\rm 2j}}{0.1} \right)   
\left( \frac{M_2/R_2}{M_\odot/R_\odot} \right)  
\\ & \times 
\left( \frac{M_2}{0.1 M_1} \right)^2
\left( \frac{\dot M_{\rm ej,rec}}{0.1  M_{\rm env} \tau^{-1}_{\rm Kep}} \right)^{-1} .
\label{eq:RatioPowers}
\end{aligned}
\end{eqnarray}
I note here that the relevant time for $\dot M_{\rm ej,rec}$ is not the ejection time itself, but rather the time it takes the ejected gas to recombine. 

The critical question is then the fraction of the jets' energy that ends in radiation. Here jets have another advantage over recombination energy. Jets can propagate and deposit most of their energy in the outer regions of the ejecta where the photon diffusion time out is shorter; see \cite{KaplanSoker2020} and \cite{Soker2020ILOTjets} for the interaction of jets with a shell and the radiation fraction. This might be significant as dust might substantially increase opacity in these transients (e.g., \citealt{GonzalezBolivaretal2023}).  

\subsection{Time scales}
\label{subsec:TimeScales}

The accretion power might exceed the Eddington limit. However, I note that the process of radiation diffusion does not reach a steady state.
To show that I consider four timescales.

I define the flow timescale to be the time during which the secondary star crosses the accretion radius under the present assumptions. This is about the time for a mass element to reach the accreting body after it crosses the accretion radius.  
Substituting the relevant variables and scaling under the present assumptions gives 
\begin{eqnarray}
\begin{aligned} 
\tau_{\rm f} \equiv \frac{R_{\rm BHL}}{v_{\rm Kep}} = 0.03 
\left( \frac{M_2}{0.1 M_1} \right)
\left( \frac{\tau_{\rm Kep}}{1 \yr} \right) \yr . 
\label{eq:TauFlow}
\end{aligned}
\end{eqnarray}
I also define the accretion timescale for the accretion disk around the secondary star, namely the viscous timescale of the accretion disk. This timescale is $\approx 100$ times the Keplerian period of the disk size. Taking the disk radius to be of the order of the secondary radius, which is adequate for a main sequence companion in a CEE, this timescale is crudely given by 
\begin{eqnarray}
\begin{aligned} 
\tau_{\rm vis} \approx & 100 \frac {2 \pi R^{3/2}_2}{\sqrt{G M_2}} = 0.03 \\ & \times
\left( \frac{M_2}{1 M_\odot} \right)^{-1/2} 
\left( \frac{R_2}{1 R_\odot} \right)^{3/2}  \yr . 
\label{eq:TauVis}
\end{aligned}
\end{eqnarray}

Consider then that the accreted mass releases energy near the surface of the secondary star. A large fraction of this energy, and even most of it, is carried by the jets on a timescale, $\tau_{\rm j,2} \simeq k_{\rm j} R_{\rm BHL} / v_{\rm j}$, where $v_{\rm j}$ is the velocity of the jets and with the parameter $k_{\rm j} > 1$ I take into account that a jet slows down as it interacts with the envelope. To power a bright transient event the jets should reach close to the surface of the envelope, and so I assume that the jets do not slow down much from their terminal velocity, which is about the escape velocity from the secondary star. I scale with $k_{\rm j} =2$. With the aid of equation (\ref{eq:Rbhl}) and $a \simeq R_1$, as the secondary is in the outer regions of the envelope, this reads for a main sequence secondary star 
\begin{eqnarray}
\begin{aligned} 
\tau_{\rm j,2} \simeq & \frac{R_{\rm BHL}}{v_{\rm j}} \simeq 0.0035
\left( \frac{k_{\rm j}}{2} \right)
\left( \frac{M_2}{0.1 M_1} \right)
\\ \times &
\left( \frac{R_1}{200 R_\odot} \right) 
\left( \frac{v_{\rm j}}{500 \km \s^{-1}} \right)^{-1} \yr. 
\label{eq:TauJ2}
\end{aligned}
\end{eqnarray}
For the photon diffusion time I take the density of the accreted gas inside $R_{\rm BHL}$ to be as that in the envelope by equation (\ref{eq:Menv}). The optical depth is then $\simeq R_{\rm BHL} \rho (R_1) \kappa$, where $\kappa$ is the opacity. The density is actually somewhat higher, increasing further the diffusion time. Under the present assumptions the diffusion time is 
\begin{eqnarray}
\begin{aligned} 
\tau_{\rm dif,2} \simeq & \frac{R_{\rm BHL}}{c} R_{\rm BHL} \rho (R_1) \kappa 
\simeq 0.2 (3- \beta) \left( \frac{\kappa}{0.4 \cm^2 \g^{-1}} \right)
\\ \times &
\left( \frac{M_2}{0.1M_1} \right)^{2}
\left( \frac{R_1}{200R_\odot} \right)^{-1} 
\left( \frac{M_{\rm env}}{1 M_\odot} \right)
  \yr . 
\label{eq:TauDif2}
\end{aligned}
\end{eqnarray}

From the above four timescales we learn the following.
The removal of energy by the jets is fast. It can bring the accretion processes to a semi-steady state (not a full steady state because the secondary star spirals-in). On the other hand, the photon diffusion time is longer than all timescales. This means that if indeed the secondary star launches jets, as I assume here, then jets carry most of the accretion energy. In that respect the Eddington limit is less relevant in reducing the accretion rate. Namely, the removal of energy by the jets makes the power carried by radiation to be sub-Eddington. The energy that the jets carry do indeed reduce the accretion rate in what is termed the negative jet feedback mechanism (for a review see  \citealt{Soker2016Rev}), as was shown numerically by \cite{Gricheneretal2021} in one-dimension and by \cite{Hilleletal2022} in three dimensions. This reduction in the accretion rate in this study enters the parameter $\eta_{\rm 2j}$. At the peak of the accretion process the power of the jets might be up to $\approx 10^2-10^3$ times the Eddington luminosity limit. This is possible for two reasons. The first is that the jets counteract accretion only along the polar directions. Indeed, no accretion occurs along the polar directions. However, accretion proceeds from the equatorial plane vicinity. The second one is that, as stated above, eventually the jets do reduce the accretion rate by depositing energy to the envelope from which accretion takes place. Namely, the highly-super-Eddington phase is temporarily one.     

\section{Implications to specific systems}
\label{sec:Systems}

\subsection{The Great Eruption of Eta Carinae}
\label{subsec:EtaCarinae}

The second peak of the Great Eruption of Eta Carinae had a rapid rise by as much as $\Delta L > 10^7 L_\odot$ that lasted few days. By equation (\ref{eq:Lrad}) a passage of the secondary inside the envelope of the primary would require the opacity to be as low as $\kappa \simeq 0.01 \cm^2 \g^{-1}$ for a secondary star of $M_2=100 M_\odot$. This is a too low opacity and a too massive secondary star for this system. Clearly a shock excited by the secondary star as it orbits through the primary stellar envelope cannot explain the Great Eruption. 

The mass loss rate in the 20 years duration of the great eruption was $\simeq 1 M_\odot \yr^{-1}$ (e.g., \citealt{Smith2008}). The companion is a massive star so that $M_2 \simeq 0.3 M_1$. The fast wind from the companion of $\simeq 3000 \km \s^{-1}$ suggests that $M_2/R_2 \ga 4 M_\odot/R_\odot$. The Keplerian period on the surface of the primary, assuming that it was the size of the periastron passage at the Great eruption, in this eccentric $e\simeq 0.9$ orbit of $5.54 \yr$ is $\simeq 0.2 \yr$ . However, I consider global properties and take here $ \tau _{\rm Kep} = 5.5 \yr$ during the Great Eruption \citep{Damineli1996}. For the envelope mass of the primary star I take $100 M_\odot$. 
Equation (\ref{eq:E2j3}) gives for the above quantities $\dot E_{\rm 2j} \approx 6 \times 10^7 L_\odot$. This energy can account for the peak luminosity of $\simeq 10^7 L_\odot$ if a fraction of $\eta_{\rm rad,2j} \simeq 0.2$ of the jets' energy is channelled to radiation.  

For the same quantities equation (\ref{eq:RatioPowers}) gives  
$\dot E_{\rm 2j} (R_1)/\dot E_{\rm rec} \approx 70$. This implies that the jets have much larger power than the recombination energy. \cite{Ivanovaetal2013a} found already that recombination energy cannot account for the Great Eruption of Eta Carinae. 
 
As the bipolar structure of the Homunculus suggests, the Great Eruption of Eta Carinae was powered by jets (e.g., \citealt{Soker2001, KashiSoker2010}). This is true whether the system did not enter a full CEE as in the binary interaction scenario for the Great Eruption, which already includes jets (e.g., \citealt{Soker2001, Soker2007}), or whether a binary system entered a CEE in the triple star scenario for the Great Eruption (e.g., \citealt{LivioPringle1998, PortegiesZwartvandenHeuvel2016, Hirai2021})\footnote{These studies of triple system progenitor of the Great Eruption  ignore jets. In any case, I find that the triple-star scenario of a merger in an unstable triple system \citep{PortegiesZwartvandenHeuvel2016, Hirai2021} has  problems. ($i$) There was a second outburst in 1890-1895 (the Lesser Eruption; \citealt{Humphreysetal1999}) that according to this scenario requires another merger. ($ii$) The present binary system shares an equatorial plane with the Homunculus nebula (e.g., \citealt{Maduraetal2012}). This is hard to explain in a non-coplanar triple system as the merger scenarios require. ($iii$) The scenario of \cite{Hirai2021} requires the presence of a dense equatorial gas that is not observed in the Homunculus. }.
In the binary model the secondary star grazed the primary stellar envelope during periastron passages and accreted mass via an accretion disk that launched the jets (e.g., \citealt{KashiSoker2010}). 
 
\subsection{V838 Mon}
\label{subsec:V838Mon}

The light curve of V838 Mon shows several peaks. A weak outburst followed by three peaks of $L \simeq 10^6 L_\odot$ \citep{Munarietal2002}. 
The rise to the first bright peak of V838 Mon lasted  for about 4 days and reached $L_{\rm peak} = 1.2 \times 10^6 L_\odot$ \citep{Tylenda2005}.  For a shock excited by a companion to explain this peak the secondary mass according to equation (\ref{eq:Lrad}) should be $>10 M_\odot$ even if the opacity is reduced to $\kappa=0.01 \cm^2 \g^{-1}$. However, the primary itself is $M_1<8 M_\odot$, and so I conclude that shock in a CEE cannot explain the bright peaks in the lightcurve of V838 Mon.  

Consider next recombination energy. 
\cite{MatsumotoMetzger2022} fit the plateau phase of V838 Mon with an ejecta mass of $M_{\rm ej} \simeq 4.2M_\odot$ and an average velocity of $860 \km \s^{-1}$. The problems with this fitting is that the ejected mass is too large for a main sequence star of mass $M_1<8 M_\odot$ and the required velocity is larger than the observed one (e.g., \citealt{Munarietal2002, Blagorodnovaetal2021}). In any case, their fitting by recombination energy does not aim to explain the 4 days rise of the first bright peak. They assume that it resulted from  cooling of the hot ejected mass. However, in a time of 4 days (the rise time of the first bright peak) gas with a velocity of $860 \km \s^{-1}$ expands to a distance of $r \simeq 3 \times 10^{13} \cm$. The optical depth is 
 $\tau \simeq \kappa M_{\rm ej} /4 \pi r^2 = 7.5 \times 10^4  (\kappa/0.1 \cm^2 \g^{-1})$. It is unlikely that at such a high optical depth the ejecta will form the rapid rise by radiative cooling. 

I keep the scaling of equation (\ref{eq:RatioPowers}) beside the last term. The effective mass ejection rate for recombination should be the mass divided by the plateau, which for V838 Mon is about 70 days. Taking $M_{\rm ej} \simeq 4.2M_\odot$ as above and the Keplerian time of the B-type progenitor to be $<0.5$~day (depending on the radius at outburst) and $M_{\rm env} \simeq 1 M_\odot$, I find from equation (\ref{eq:RatioPowers}) 
 $\dot E_{\rm 2j} (R_1)/\dot E_{\rm rec} \approx 100$. 
For these quantities equation (\ref{eq:E2j3}) gives (with the other parameters as in the scaling) $\dot E_{2j} \approx 10^8 L_\odot$, implying that the jets can account for the kinetic energy of the ejecta and the radiation. 

Before concluding I note that the outburst of V838 Mon could have resulted from the disruption of a low-mass pre-main sequence star on the primary star rather than the onset of a CEE (\citealt{SokerTylenda2003,  TylendaSoker2006}). I still claim it was powered by jets, but in that case jets that the primary star launched as it accreted the disrupted secondary mass via an accretion disk.  

I conclude that V838 Mon was most likely powered by jets, in particular the first bright peak. Indeed, the bipolar structure of its inner ejecta that was established recently by \cite{Mobeenetal2023} (for earlier hints see  \citealt{Chesneauetal2014, Kaminskietal2021, Mobeenetal2021}) and the point-symmetry of the clumps they find strongly suggests that jets were the main powering source of this ILOT. The shock that the companion excites in the primary envelope and recombination energy add some small fraction to the powering and radiation.

\subsection{V1309 Scorpii}
\label{subsec:V1309 corpii}

V1309 Scorpii  had a roughly exponential rise in brightness during outburst, rising by a factor of about 100 in the last 10 days to $L=3 \times 10^{4} L_\odot$ \citep{Tylendaetal2011}. \cite{Nandezetal2014} build a model for V1309 where the primary and secondary masses are $M_1=1.52 M_\odot$ and $M_2=0.16 M_\odot$, respectively. Unless the companion is more massive and opacity much lower than $0.1 \cm^2 \g^{-1}$, I find from equation (\ref{eq:Lrad}) that shock at CEE formation cannot explain the peak of this outburst. 

I examine powering by recombination. The lightcurve of V1309 Scorpii has a faint peak (e.g., compared with V838 Mon), and the rise to the maximum is exponential rather than with a steep step. \cite{MatsumotoMetzger2022} fit the peak with an ejecta mass of $M_{\rm ej} \simeq 0.032 M_\odot$ and an average velocity of $550 \km \s^{-1}$. Although the ejected mass is easily accommodate by a progenitor of $M_1 \simeq 1 M_\odot$, the required velocity is much larger than the observed velocity of the ejecta of V1309 Scorpii which is $\simeq 150 \km \s^{-1}$ (\citealt{Tylendaetal2011, Blagorodnovaetal2021}). Another problem is that the light curve that results from recombination energy as \cite{MatsumotoMetzger2022} calculate has a rise time about equal to decline time. In V1309 Scorpii the rise time is much shorter than the decline time \citep{Tylendaetal2011}. 

With a primary envelope mass of $M_{\rm env} = 1 M_\odot$, a Keplerian time of $\tau_{\rm Kep}=0.1$~day, ejected mass of $M_{\rm ej} = 0.032 M_\odot$ as above, and a plateau time of $\simeq 20$~days, which is the relevant time for the recombination timescale and which gives $\dot M_{\rm ej,rec} \simeq 0.6 M_\odot \yr^{-1}$, I find from equation (\ref{eq:RatioPowers}) that 
 $\dot E_{\rm 2j} (R_1)/\dot E_{\rm rec} \approx 2000$. Clearly accretion power at the peak can dominate the recombination power. 
 
In seems that V1309 Scorpii was also powered by jets with a lesser contribution from recombination. I predict that its ejecta has a bipolar morphology. Observations in about a decade or two might resolve the structure of the ejecta (for a distance of $3 \kpc$ and an expansion velocity of $150 \km \s^{-1}$; \citealt{Tylendaetal2011}).

\subsection{Progenitors of planetary nebulae}
\label{subsec:PNe}

I consider the onset of CEE of RGB and AGB stars with main sequence secondary stars, i.e., progenitor systems of planetary nebulae.  
\cite{SokerKashi2012} suggested that some bipolar PNe formation involve one or more ILOT events. They could only estimate the luminosity of these speculated ILOTs, and give the range of $L_{\rm PN,ILOT} \simeq 3 \times 10^4 L_\odot - 10^7 L_\odot$. More reasonable range would be  $L_{\rm PN,ILOT}  \simeq 10^5 L_\odot - 10^6 L_\odot$. The formation of a bright ILOT during mass ejection of a bipolar PN requires a relatively massive secondary star, i.e., $M_2 \ga 0.5 M_\odot$ and that its jets shape the large lobes.

Simulations show that without the inclusion of the recombination energy of the unbound gas the mass ejection rate during CEE of RGB, AGB and RSG stars is $<M_{\rm env}/\tau_{\rm Kep}$ (e.g., \citealt{NandezIvanova2016, Krameretal2020, Reichardtetal2020, Sandetal2020, GlanzPerets2021, GonzalezBolivaretal2022, Lauetal2022}; see \citealt{RoepkeDeMarco2023} for a review). 
High mass ejection rates of $\approx M_{\rm env}/\tau_{\rm Kep}$ can also be achieved without recombination energy if the secondary star is massive, $M_2 \ga 0.5 M_1$,  (e.g., \citealt{RickerTaam2012}). 

\cite{Sandetal2020} conduct CEE simulations with an AGB primary star. They find that the mass ejection rates can reach values close to $M_{\rm env} / \tau_{\rm Kep}$ only when the recombination energy is channelled to mass ejection\footnote{I do not refer to the very early mass ejection rates that their graphs show at the beginning of the simulations as these seem numerical adjustments.}. 
In that case of course the recombination energy does not contribute much to the lightcurve of the event. 

Overall, when most of the recombination energy goes to radiation, the scaling of equation (\ref{eq:RatioPowers}) is appropriate for binary progenitors of planetary nebulae. For a more massive secondary star with $M_2 > 0.1 M_1$ the mass ejection rate might be larger, but the inequality $\dot E_{\rm 2j} (R_1) > \dot E_{\rm rec}$, or even $\dot E_{\rm 2j} (R_1) \gg \dot E_{\rm rec}$, hold. Again, photon diffusion time out from the interaction zone of the jets with the shell (ejecta), $t_{\rm diff,ej}$,  is expected to be shorter than the recombination timescale, making the rise time shorter for jets' powering than for recombination. 
Overall, I conclude that bright peaks with short rise time during CEE formation of planetary nebula progenitors are likely to be powered by jets. 

As for the peak luminosity itself at CEE formation, the scaling of equation (\ref{eq:E2j3}) is appropriate, but for bipolar planetary nebulae with a main sequence companion I would expect $M_2 \ga 0.3 M_1$. Not all the jets' energy ends in radiation, but when the companion is in the outer zone of the envelope, a large fraction $\eta_{\rm rad,2j}$ might end in radiation. Overall, I scale the jet-powered peak radiation at CEE formation of planetary nebula progenitors by the crude estimate 
\begin{eqnarray}
\begin{aligned} 
& L_{\rm CE,PN}  \simeq 3 \times 10^5 
(3-\beta)  
\left( \frac{\eta_{\rm rad,2j}}{0.3} \right)
\left( \frac{\eta_{\rm 2j}}{0.1} \right)   
\\ & \times 
\left( \frac{M_2}{0.3 M_1} \right)^2
\left( \frac{M_2/R_2}{M_\odot/R_\odot} \right)  
\left( \frac{M_{\rm env}}{1 M_\odot} \right)
\left( \frac{\tau_{\rm Kep}}{1 \yr} \right)^{-1} L_\odot
\label{eq:LradPNe}
\end{aligned}
\end{eqnarray}
This is much brighter than the emission due the shock that the orbital motion of the companion excites in the envelope (equation \ref{eq:Lrad}). 

The peak in the light curve that the jets power might last for a time much  shorter than the Keplerian time. 
The emission at later times after the companion plunge into the envelope will come from both jets and recombination energy, or only from recombination energy. The peak emission at this phase will be longer and fainter that the peak at CEE formation.   

I summarize the relevant properties of the three ILOTs and the CEE formation by planetary nebula progenitors in Table \ref{Table}. 
\begin{table*}[]
\caption{Binary properties of studeid ILOTs}
\centering
\begin{tabular}{|l|c|c|c|c|c|c|c|}
\hline
Object & $M_1$ & $M_2$ & $\dot M_{\rm ej,rec}$ & $M_{\rm env}$ & $\tau_{\rm Kep}$ & $\dot E_{\rm 2j}$  & ${\dot E_{\rm 2j}}/{\dot E_{\rm rec}}$ \\ 
& $M_\odot$ & $M_\odot$ & $M_\odot \yr^{-1}$ & $M_\odot$ & year & $L_\odot$ & \\
\hline
GE Eta Car & $\simeq 120-200$ & $\simeq 30-90$ & $\simeq 1$ & $=100$ & $=5.5$ & $\approx 10^8$ & $\approx 70$ \\
V838 Mon      & $=6$ & $\simeq 0.5$ & $\approx 10$ & $\simeq 1$ & $\simeq 10^{-3}$ & $\approx 10^8$ & $\approx 100$ \\
V1309 Scorpii & $= 1.5$ & $\simeq 0.15$ & $\approx 0.6$ & $\simeq 1$& $3 \times 10^{-4}$ & $\approx 10^8$ & $\approx 2000$ \\
CEE-PNe       & $=1-5$ & $\simeq 0.3-1$ & $[\approx 0.1]^\star$ & $\simeq 1$ & $\simeq 1$ & $\approx 40$ & $[\approx 10^6]^\star$ \\
\hline
\end{tabular} \\
\vspace{0.1cm}
\begin{minipage}{12.7cm}
{The binary properties (columns 2-6) that I use in this study to calculate the jets' power (next to last column by equation \ref{eq:E2j3}) and its ratio to the recombination power (last column by equation \ref{eq:RatioPowers}). The input parameters from left to right are the name of the system, the primary stellar mass, the secondary stellar mass, the recombination rate of the ejected gas if recombination powers the plateau of the lightcurve, the envelope mass of the primary star, and the Keplerian period on the surface of the primary star. The sources to the different values are from papers cited in the text. The input parameters of the different systems can be within uncertainties of up to few tens of percent ($=$), up to a factor of a few ($\simeq$), or up to about an order of magnitude ($\approx$). The derived quantities are uncertain by about an order of magnitude.  
Abbreviation: 
CEE-PNe: CEE formation by planetary nebula progenitors;
GE Eta Car: The Great Eruption of Eta Carinae. 
$\star$ In the case of CEE formation by planetary nebula progenitors there are no direct observations of any ILOT with an estimated ejecta recombination rate. I therefore take $\dot M_{\rm ej,rec} \approx 0.1 M_\odot \yr^{-1}$ as a demonstrative example.}
\end{minipage}
\label{Table}
\end{table*}
\normalsize
\section{Summary}
\label{sec:Summary}

I compared with each other and with observations three energy sources to power ILOTs (LRNe) at CEE formation (including GEE formation). 
I presented the power of the energy sources in terms of global properties of the binary system for the shock that the secondary star excites in the envelope of the primary star (equation \ref{eq:Lrad}), for the power of the jets that the secondary star launches (equation \ref{eq:E2j3}), and for the recombination energy (equation \ref{eq:Erec}). Equation (\ref{eq:RatioPowers}) gives the ratio of jets' power to recombination power. 
In comparing to the properties of emission powered by the recombination energy I used also the results of  \cite{Ivanovaetal2013a} and \cite{MatsumotoMetzger2022}. 

I compared these energy sources with three observed ILOTs, the Great Eruption of Eta Carinae in section \ref{subsec:EtaCarinae}, with V838 Mon in section \ref{subsec:V838Mon}, and with V1309 Scorpii in section \ref{subsec:V1309 corpii}. In all three cases the shock that the secondary star excites in the envelope falls short of explaining the peak luminosity in the lightcurve. In all three cases recombination powering yields a  rise time to peak luminosity that is too long to explain the observed rapid rise to the first bright peak. In contrast, I find that the energy that the jets can supply explains the rapid rise and peak luminosity in all three ILOTs. 
I predict that the ejecta (nebula) of V1309 Scorpii will be observed in a decade or two to be bipolar. I summarised the relevant properties in table \ref{Table}.  

Although the claim for jet-powering of these ILOTs have been made in the past (see section \ref{sec:Systems}), this study for the first time systematically compares the three energy sources with each other and with these ILOTs. This study also presents the jets' power by the global properties of the primary star, i.e., the Keplerian period on the surface of the primary star and its total envelope mass (rather than the local properties of density and velocity). Most of all, this study aims at the CEE formation phase. 

In section \ref{subsec:PNe} I applied the results to CEE (or GEE) formation phase during the ejection of planetary nebulae. I presented the scaled equation (\ref{eq:LradPNe}) for the expected peak luminosity when a main sequence companion enters the envelope of an RGB star or an AGB star. This peak luminosity is compatible with the range of estimated luminosities of the ILOTs that \cite{SokerKashi2012} speculated to have occurred at the formation of four bipolar planetary nebulae. I concluded that only jets can power bright ILOTs (LRNe) at CEE formation. 

This study adds to the accumulating evidences, observationally and theoretically, that jets serve as the main powering source of most ILOTs (including LRNe and luminous blue variable major eruptions).

\section*{Acknowledgements}
I thank Amit Kashi  and an anonymous referee for useful comments. This research was supported by a grant from the Pazy Research Foundation.


\end{document}